\documentclass[pra,twocolumn]{revtex4}
\usepackage{amsmath}
\usepackage{graphicx}
\usepackage{bm}

\begin{document}
\title{Coherent Logic Gate for Light Pulses based on Storage in a Bose-Einstein Condensate}
\author{Christoph Vo}
\author{Stefan Riedl}
\altaffiliation{Present address: National Institute of Standards and Technology, 325 Broadway, Boulder, Colorado 80305, USA}
\author{Simon Baur}
\author{Gerhard Rempe}
\author{Stephan D\"{u}rr}
\affiliation{Max-Planck-Institut f\"{u}r Quantenoptik, Hans-Kopfermann-Stra{\ss}e 1, 85748 Garching, Germany}

\begin{abstract}
A classical logic gate connecting input and output light pulses is demonstrated. The gate operation is based on three steps: First, two incoming light pulses are stored in a Bose-Einstein condensate, second, atomic four-wave mixing generates a new matter wave, and third, the light pulses are retrieved. In the presence of the new matter wave, the retrieval generates a new optical wave. The latter will only be generated if both input light pulses are applied, thus realizing an AND gate. Finally, we show that the gate operation is phase coherent, an essential prerequisite for a quantum logic gate.
\end{abstract}

\maketitle

Single photons are well suited for quantum communication over long distances. To perform quantum information processing with single photons, however, one must find a physical process in which a single photon drastically alters some property of another single photon. This is a major challenge because in traditional nonlinear optical media the nonlinearities are much too weak to generate an appreciable effect on the single-photon level. Several techniques for addressing this problem have been proposed and are being pursued experimentally, namely the use of atoms in optical resonators \cite{Turchette:95, duan:04, Imamoglu:97}, the use of additional light to drive Raman transitions in atoms \cite{Imamoglu:97, Lukin:01, Bajcsy:09}, and the use of the dipole-dipole interaction between Rydberg atoms \cite{Gorshkov:11, Dudin:12, Peyronel:12}.

Here we present a first experiment that explores the avenue of generating a logic gate for classical light pulses by temporarily converting the light pulses into atomic excitations in a Bose-Einstein condensate (BEC) and using $s$-wave collisions between pairs of ground-state atoms. These collisions are responsible for the appearance of the nonlinear term in the Gross-Pitaevskii (GP) equation. In the context of quantum information processing, they have been used to generate massive entanglement between many atoms \cite{mandel:03} but not to generate a logic gate for light pulses. In addition, we demonstrate that the gate operation is phase coherent, an essential prerequisite for a quantum logic gate.

We use a geometry in which the nonlinearity of the GP equation creates a new atomic momentum component by four-wave mixing (FWM) of matter waves \cite{Goldstein:95, Trippenbach:98, deng:99, Trippenbach:00} involving two spin states \cite{Goldstein:99, pertot:10}. Upon mapping the new atomic momentum component back onto light, it creates population in a new optical momentum component. This light emission process is accompanied by Raman amplification of matter waves (AMW) \cite{Schneble:04, Cola:04}. The light emitted during Raman AMW and the phase coherence of this light have never been studied experimentally; despite related work in atomic FWM \cite{Vogels:03}, Rayleigh AMW \cite{Kozuma:99, Inouye:99:Phase}, and superradiant light scattering \cite{Inouye:99:Superradiance, Inouye:00}.

\begin{figure}[t!]
\includegraphics[width=0.8\columnwidth]{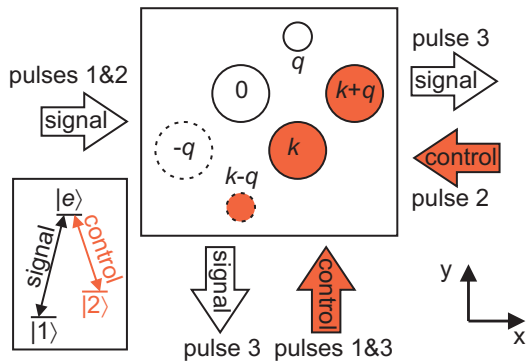}
\caption{
\label{fig-scheme}
(Color online) \textbf{Scheme of the experimental procedure.} Two Raman pulses are applied, preparing three BECs with momenta $0$, $\bm k$, and $\bm k+\bm q$. During a subsequent dark time, atomic FWM creates a BEC with momentum $\bm q$. Arrows show the propagation directions of the light beams. Circles represent atomic momentum components. The internal state is color coded: white=$|1\rangle$, red=$|2\rangle$. A third light pulse with only control light applied retrieves the signal light. The retrieved light will have a component propagating downward only if signal light is applied during both Raman pulses, thus realizing an AND gate. A modified state preparation can additionally populate the momentum component $-\bm q$ so that atomic FWM also populates $\bm k-\bm q$. This extended scheme, in which the dashed circles are populated, generates a time-dependent interference pattern in the retrieved light.
}
\end{figure}

A scheme of our experiment is shown in Fig.\ \ref{fig-scheme}. The $^{87}$Rb hyperfine states $|1\rangle=|F=1\rangle$ and $|2\rangle=|F=2\rangle$ of the $5S_{1/2}$ ground state together with the $|e\rangle=|5P_{1/2}\rangle$ excited state, each with $m_F=-1$, form a $\Lambda$ scheme in which Raman transitions are driven. More precisely, the Raman light fields are tuned to the two-photon resonance with the single-photon detuning chosen exactly midway between the $|1'\rangle=|F'=1\rangle$ and $|2'\rangle=|F'=2\rangle$ components of the excited state $|e\rangle$.

Initially, a BEC with $N\sim1.5\times 10^6$ atoms is prepared in an optical dipole trap with measured trap frequencies of $(\omega_x,\omega_y,\omega_z)=2\pi\times(70,20,20)$ Hz, in internal state $|1\rangle$, and at zero momentum, which we denote as $|1,0\rangle$. A magnetic hold field of $\sim 1$ G is applied along the $z$-axis (orthogonal to the plane shown in Fig.\ \ref{fig-scheme}) to preserve the atomic spin orientation. All light fields applied in the experiment drive $\pi$ transitions.

After preparation, the BEC is illuminated by a Raman pulse, consisting of signal light with wave vector $\bm k_s$, propagating rightward in Fig.\ \ref{fig-scheme}, and control light with wave vector $\bm k_{c1}$, propagating upward. Signal light is absorbed and coherently stored in the atomic state $|2,\bm k\rangle$ with internal state $|2\rangle$ and wave vector $\bm k=\bm k_s-\bm k_{c1}$. The pulse area of the Raman pulse is chosen such that $\sim 1/3$ of the atomic population is transferred to state $|2,\bm k\rangle$.

Immediately thereafter, a second Raman pulse is applied with the direction of the signal light as before, but now with the control light propagating leftward with wave vector $\bm k_{c2}$. During this pulse, signal photons are absorbed and stored in state $|2,\bm k+\bm q\rangle$ with $\bm q=\bm k_s-\bm k_{c2}-\bm k$ \cite{supplement:model:experiment}. This Raman pulse has a duration of $\sim 100$ $\mu$s. The pulse area of $\sim \pi/2$ yields equal populations of states $|1,0\rangle$, $|2,\bm k\rangle$, and $|2,\bm k+\bm q\rangle$.

In principle, the second Raman pulse could simultaneously drive a second process in which population is transferred from state $|2,\bm k\rangle$ to state $|1,-\bm q\rangle$. In practice, however, the nonzero initial momentum creates a Doppler shift for the resonance frequency of this process. By fine tuning the two-photon detuning we resonantly drive the $|1,0\rangle \to |2,\bm k+\bm q\rangle$ processes and drastically suppress the $|2,\bm k\rangle \to |1,-\bm q\rangle$ process. The pulse is long enough to make interaction-time broadening small compared to the energy splitting between the two resonances frequencies.

During the following dark time, with duration $t_\mathrm{FWM}$, atomic FWM with two internal states populates the state $|1,\bm q\rangle$. The FWM can be understood intuitively as an atomic scattering process. An atom in state $|1,0\rangle$ collides with an atom in state $|2,\bm k+\bm q\rangle$. The existing BEC in state $|2,\bm k\rangle$ creates bosonic enhancement for one atom to emerge in this state. Conservation of momentum and of the internal-state energy makes the other atom appear in state $|1,\bm q\rangle$. In addition, conservation of kinetic energy requires $\bm k\cdot \bm q\sim 0$ \cite{Trippenbach:98, deng:99}. Note that the FWM process in our experiment is clearly distinguishable from spin exchange, unlike the only previous experiment on atomic FWM with two internal states \cite{pertot:10}. The Bose-enhanced creation of other atomic momentum components $|1,n \bm q\rangle$ and $|2,\bm k+n\bm q\rangle$ with integer $n$ would conserve momentum but not kinetic energy and is therefore negligible. The FWM occurs inside the optical dipole trap in order to avoid a slowdown of the FWM due to the reduced density in a mean-field driven expansion.

\begin{figure}[t!]
\includegraphics[width=0.85\columnwidth]{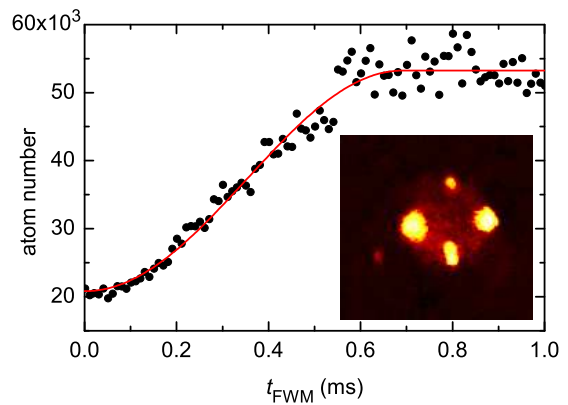}
\caption{
\label{fig-absorption}
(Color online) \textbf{Atomic four-wave mixing.} The atom number $N_q$ in the atomic momentum component $\bm q$ is extracted from time-of-flight absorption images. $N_q$ depends on the four-wave mixing time. At short times, this dependence is quadratic. Inset: An image oriented as Fig.\ \ref{fig-scheme}, taken for $t_\mathrm{FWM}=1.8$ ms without applying depletion light.
}
\end{figure}

To test whether atomic FWM occurs in our experiment, we study time-of-flight absorption images. The inset in Fig.\ \ref{fig-absorption} shows such an image. Four atomic momentum components are clearly distinguishable. The additional tiny signal in the bottom left corner shows that the process populating state $|1,-\bm q\rangle$ during the second Raman pulse is suppressed drastically but not completely.

To confirm that the atom number $N_q$ in the momentum component $|1,\bm q\rangle$ is actually generated by FWM, we study its temporal growth. To avoid FWM during the time of flight, we abort the FWM by applying an 8 $\mu$s pulse of depletion light \cite{supplement:model:experiment}, 65 MHz blue detuned from the $|2\rangle \leftrightarrow |1'\rangle$ transition, immediately before release from the dipole trap. The atom number $N_q$ extracted from time-of-flight images taken after applying depletion light is shown in Fig.\ \ref{fig-absorption}. For short times, $N_q$ displays a quadratic growth with time, as expected for FWM. For longer times, $N_q$ saturates because the different momentum components no longer overlap spatially. The line shows a fit to the data, where the observed timescales for the initial growth and for the saturation agree fairly well with theory \cite{supplement:model:experiment}.

After confirming that $N_q$ is actually generated by FWM, we now map the atomic states back onto the light field. To this end, the BEC in the trap is illuminated by a third light pulse, the retrieval pulse, during which a control beam propagating upward is applied, whereas no signal light is applied. We use a detuning of 300 MHz red from the $|2\rangle \leftrightarrow |2'\rangle$ transition because we find experimentally that this maximizes the retrieved photon number propagating downward. Using control light with an intensity of roughly 100 mW/cm$^2$, we find that the retrieved light emerges in less than 5 $\mu$s. Each atom in internal state $|2\rangle$ is transferred back into internal state $|1\rangle$ in a Raman process under emission of a signal photon. These Raman processes are bosonically stimulated by the two BECs in states $|1,0\rangle$ and $|1,\bm q\rangle$. The stimulated growth of atomic population in these two BECs is called Raman AMW. Along with the population growth in states $|1,0\rangle$ and $|1,\bm q\rangle$, signal light with two momentum components is emitted, one propagating downward, the other rightward.

\begin{figure}[t!]
\includegraphics[width=0.75\columnwidth]{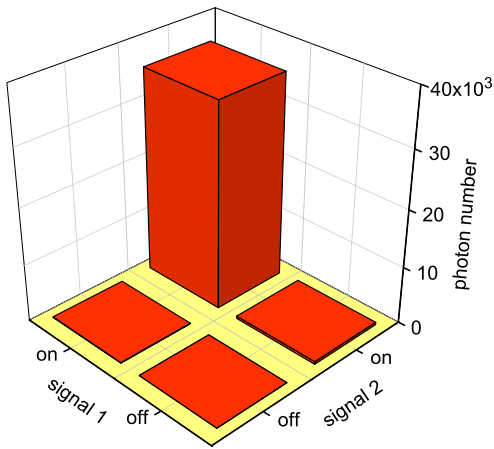}
\caption{
\label{fig-gate}
(Color online) \textbf{Logic gate.} The retrieved photon number propagating downward is shown for four different experimental settings, in which the signal beams during Raman pulses 1 and 2 are turned on or off independently. Downward propagating light will only be retrieved if both signal beams are on. This demonstrates an AND gate for classical light pulses based on storage in a BEC and FWM of matter waves.
}
\end{figure}

We concentrate on the downward propagating component with wave vector $\bm k_s-\bm q$. It is created when atoms are transferred from  state $|2,\bm k\rangle$ to $|1,\bm q\rangle$. This will only be possible if atomic FWM occurs because otherwise $N_q=0$. FWM, in turn, will occur only if the signal light is applied during both Raman pulses. We use an electron-multiplying charge-coupled device (EMCCD) camera to measure the photon number propagating downward. An iris diaphragm is placed in an intermediate imaging plane to suppress stray light from the control beam. Results for $t_\mathrm{FWM}=0.4$ ms are shown in Fig.\ \ref{fig-gate}. They clearly demonstrate an AND gate for the two classical signal light pulses.

The rightward propagating component of the retrieved light is of less interest here. Its existence does not rely on atomic FWM. If the second pulse were omitted completely, there would still be the retrieval of rightward propagating light, well known from experiments on electromagnetically induced transparency (EIT) \cite{fleischhauer:05}. The photon number retrieved in this beam, however, does depend on whether FWM occurred because there is competition between the processes retrieving light propagating downward and rightward.

When considering the perspectives for scaling this gate down to the single-photon level in order to obtain a quantum logic gate, it is crucial whether the gate operation is phase coherent. We will now show experimentally that this is the case for the gate demonstrated here.

From a theoretical point of view \cite{Cola:04}, Raman AMW is analogous to a usual stimulated Raman process, except that the emission is bosonically stimulated not by application of a second laser beam but by the presence of a second BEC. Effectively the role of the second laser beam and the second BEC are exchanged. In a usual stimulated Raman process, the atoms are transferred into the initially empty state in a phase coherent way, with the relative phase of the two applied laser beams determining the phase of the transferred atomic amplitude. Similarly, we expect that Raman AMW generates light in a phase coherent way, with the relative phase of the two BECs determining the phase of the emitted light.

To test experimentally whether the emitted light is phase coherent, we shorten the second Raman pulse to $\sim 35$ $\mu$s at correspondingly higher light intensities. Interaction-time broadening now makes the Doppler shift for the transfer of population into state $|1,-\bm q\rangle$ irrelevant. We choose pulse areas of $\sim \pi/2$ for both Raman pulses to create four equally populated atomic momentum components $|1,0\rangle$, $|1,-\bm q\rangle$, $|2,\bm k\rangle$, and $|2,\bm k+\bm q\rangle$. Subsequently, atomic FWM populates the states $|1,\bm q\rangle$ and $|2,\bm k-\bm q\rangle$, see Fig.\ \ref{fig-scheme}. We refer to this as the extended scheme because it features six atomic momentum components, instead of four. The retrieval pulse, applied as before, again will generate downward propagating signal light only if atomic FWM occurs. But now, two pathways contribute to this signal. The light can be generated by transfer of an atom either from state $|2,\bm k\rangle$ to $|1,\bm q\rangle$ or from $|2,\bm k-\bm q\rangle$ to $|1,0\rangle$.

\begin{figure}[t!]
\includegraphics[width=0.85\columnwidth]{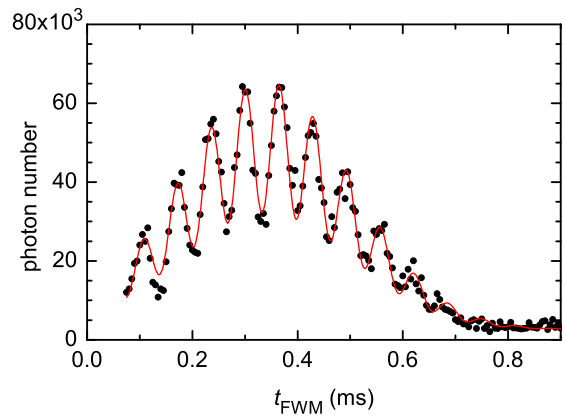}
\caption{
\label{fig-oscillation}
(Color online) \textbf{Phase coherence.} The retrieved photon number propagating downward exhibits a sinusoidal oscillation as a function of the FWM time. The best-fit value for the oscillation frequency is 15.4 kHz. The oscillation occurs because two pathways contribute to this signal in the extended scheme with six momentum components.
}
\end{figure}

The retrieved photon number propagating downward is shown in Fig.\ \ref{fig-oscillation} as a function of $t_\mathrm{FWM}$. It clearly shows a sinusoidal oscillation. To understand the physical origin of this oscillation, we note that during $t_\mathrm{FWM}$, the atomic components that contribute to the first pathway, $|2,\bm k\rangle \to |1,\bm q\rangle$, differ in kinetic energy by $\Delta E_1=\hbar^2(k^2-q^2)/2m$, whereas the components of the second pathway, $|2,\bm k-\bm q\rangle \to |1,0\rangle$, differ by $\Delta E_2=\hbar^2(\bm k-\bm q)^2/2m$. Here $m$ denotes the atomic mass. During $t_\mathrm{FWM}$, these pairs of BECs thus accumulate different relative phases. Upon retrieval, these relative phases of the pairs of BECs are mapped onto the phases of the retrieved signal light, as discussed above. Eventually, the amplitudes associated with the two pathways create an interference signal on the detector. An independent measurement of our beam geometry yields $q^2=2.08 k_s^2$ and $\bm q\cdot \bm k=-0.037 k_s^2$. With $2\pi/k_s=794.979$ nm we expect an angular frequency $\omega=(\Delta E_2-\Delta E_1)/\hbar=2\pi \times 15.4$ kHz for the oscillation. If the beams pointed exactly along the coordinate axes, then we would expect $\omega=4E_\mathrm{rec}/\hbar$, where $E_\mathrm{rec}=\hbar^2 k_s^2/2m$ is the recoil energy of the signal light.

Two effects contribute to the observed envelope of the oscillation \cite{supplement:model:experiment}. First, Raman AMW requires spatial overlap of the atomic momentum components. For long times, such overlap is lost and no directed retrieval is obtained. The timescale seen for this effect in Fig.\ \ref{fig-oscillation} is similar to the timescale of saturation in Fig.\ \ref{fig-absorption}, as expected. Second, retrieval of downward propagating light requires FWM to occur, so that an increase at short times, as seen in Fig.\ \ref{fig-absorption}, is also seen in the envelope in Fig.\ \ref{fig-oscillation}.

For simplicity, we fit a sinusoid with a Gaussian envelope and a constant visibility $V$ to the data in Fig.\ \ref{fig-oscillation}. This yields best-fit values of $\omega/2\pi=15.4\pm0.1$ kHz and $V=0.35\pm0.02$. The fact that $V$ is not perfect is theoretically expected \cite{supplement:model:experiment}. The agreement with the above expectation for $\omega$ is excellent. This oscillation is not seen in time-of-flight absorption images, either in Fig.\ \ref{fig-absorption} or in similar data that we took for the extended scheme (not shown here). This demonstrates that the retrieval of the light is necessary to make the oscillation appear.

The observed oscillation proves that the complete gate operation is phase coherent. This includes all physical processes involved in the gate, namely Raman pulses 1 and 2, atomic FWM, and light emission during Raman AMW. The observed phase coherence is a crucial ingredient for a possible extension to a photon-photon quantum logic gate.

It should be noted that an extension of our scheme to the single-photon level is challenging. One possible problem is that smaller photon numbers slow down the atomic FWM so that the spatial overlap of the matter waves might end before achieving enough population transfer. Another possible problem is that the bosonic stimulation caused by a single atom in driving either atomic FWM or Raman AMW is weak so that competing processes might cause problems, e.g., isotropic $s$-wave scattering or spontaneous photon emission. Working out strategies to solve these problems is beyond the scope of the present paper.

On a more general level, our experiment demonstrates that storage and retrieval of light combined with the nonlinearity of the GP equation can be used to create a phase-coherent gate for two classical light pulses. If the geometry were altered to make all laser beams copropagating, then the nonlinearity of the GP equation would generate a conditional phase shift instead of atomic FWM. A recent theoretical analysis of the single-photon version of that scheme resulted in a detailed proposal for a photon-photon quantum-logic gate \cite{Rispe:11}.

We thank M. Lettner for discussions and D. Fauser for assistance with the experiment. This work was supported by the German Excellence Initiative through the Nanosystems Initiative Munich and by the Deutsche Forschungsgemeinschaft through SFB 631.

\section*{APPENDIX}

\section{Dynamics of Four-Wave Mixing}

\subsection{Initial Growth of $N_q(t)$}

The growth of $N_q(t)$ at short times can be estimated as follows. Let $N_0$, $N_q$, $N_k$, and $N_{k+q}$ denote the atom numbers with the various momenta, $a_{12}$ the interspecies scattering length, and $g_{12}=4\pi\hbar^2 a_{12}/m$. We write the single-particle wave-function of a Thomas-Fermi parabola as \begin{eqnarray}
\label{u}
u(\bm x) = \frac1{\sqrt {\mathcal V}} \left(1-\sum_{i=1}^3 \frac{x_i^2}{R_i^2} \right)^{1/2}
,\end{eqnarray}
where this is real and $u(\bm x)=0$ otherwise. The normalization condition $\int d^3x |u(\bm x)|^2=1$ yields
\begin{eqnarray}
\label{V}
\mathcal V=\frac{8\pi}{15} R_x R_y R_z
.\end{eqnarray}
$\mathcal V$ has the dimension of a volume and expresses the ratio of particle number over peak density. From the measured trap frequencies and the measured initial total particle number, the Thomas-Fermi radii in our experiment are estimated to be $(R_x,R_y,R_z)=(8,27,27)$ $\mu$m.

Based on coupled Gross-Pitaevskii equations \cite{pertot:10:S}, one can generalize Eq.\ (17) of Ref.\ \cite{Trippenbach:00:S} to the case of FWM with two internal states. This yields the following result for the condensate wave function of the component $|1,\bm q\rangle$ at short times
\begin{eqnarray}
\psi_q(\bm x,t) = \frac{g_{12}t}{i\hbar} u(\bm x) |u(\bm x)|^2 \sqrt{N_0(0) N_k(0) N_{k+q}(0)}
.\end{eqnarray}
This is normalized such that $N_q(t)=\int d^3x \linebreak[1] |\psi_q(\bm x,t)|^2$. Using
\begin{eqnarray}
\int d^3x |u(\bm x)|^6 = \frac{8}{21\mathcal V^2}
\end{eqnarray}
we obtain
\begin{eqnarray}
\label{N-q}
N_q(t)
= \left(\frac{t}{\tau}\right)^2
,\end{eqnarray}
where we introduced the timescale for the initial growth
\begin{eqnarray}
\label{tau}
\tau
= \frac{\hbar \mathcal V}{g_{12}} \sqrt{\frac{21}{8N_0(0) N_k(0) N_{k+q}(0)}}
.\end{eqnarray}
For $[N_0(0)N_k(0)N_{k+q}(0)]^{1/3}=2.8\times10^5$ and $a_{12}=98.4$ Bohr radii \cite{Kokkelmans:S}, Eq.\ \eqref{tau} predicts $\tau=2.1$ $\mu$s.

\subsection{Fitting Function for Fig.\ \ref{fig-absorption}}
\label{supp-sec-fitting-function}

At longer times, FWM can make the atomic population oscillate back and forth between different momentum components, at least in principle \cite{Trippenbach:00:S}. In our experiment, however, the FWM dynamics come to an end because the different momentum components spatially separate as time progresses. The FWM dynamics is terminated at a time when the fraction of the atomic population that has been transferred is still small and no oscillations have occurred yet.

Using two coarse approximations, we now derive an analytic expression for $N_q(t)$ for all times. To represent the decreasing spatial overlap of the clouds at long times, we take the different center-of-mass momenta of the momentum components into account. However, we neglect a possible time dependence of shape, radii, and particle numbers of the three initially populated Thomas-Fermi parabolae. Hence, the time dependence of the condensate wave functions for $\alpha\in\{0,k,k+q\}$ is approximated as
\begin{eqnarray}
\label{psi-alpha}
\psi_\alpha(\bm x,t)=
\sqrt{N_\alpha(0)} \; u\left(\bm x-\frac{\hbar \bm k_\alpha}{m} t\right)
\end{eqnarray}
with $\bm k_\alpha=0$, $\bm k$, $\bm k+\bm q$ for $\alpha=0$, $k$, $k+q$, respectively. In our experiment, the FWM transfers only a small fraction of the population, all relevant scattering lengths are almost identical, and the timescale for the clouds to separate spatially is much shorter than the trapping period. Hence these approximations seem reasonable.

Furthermore, we simplify Eq.\ (15) of Ref.\ \cite{Trippenbach:00:S} by neglecting mean-field energies, external potential, and dispersion of the newly generated wave packet. This yields
\begin{eqnarray}
\left(\partial_t + \frac{\hbar \bm q}m \cdot \nabla \right) \psi_q
= \frac{g_{12}}{i\hbar} \psi_0 \psi_k^* \psi_{k+q}
.\end{eqnarray}
This is our first coarse approximation. It drastically simplifies the problem. For $g_{12}=0$, the solution would be a classical drift of the initial wave packet $\psi_q(\bm x,t)=\psi_q(\bm x-\hbar \bm q t /m,0)$, much like Eq.\ \eqref{psi-alpha}. For $g_{12}\neq0$, each drift trajectory $\bm x(t)=\bm x(0)+\hbar \bm q t /m$ accumulates an amplitude due to FWM but there is no crosstalk between different trajectories because we neglected the dispersion of the wave packet. The initial condition is $\psi_q(\bm x,0)=0$.

Considering the trajectory that originates at the cloud center $\bm x(0)=0$, we obtain
\begin{eqnarray}
\lefteqn{
\psi_q\left(\frac{\hbar \bm q}m t ,t \right)
}\nonumber \\
&=& \frac{g_{12}}{i\hbar} \int_0^t dt_3 \psi_0(\bm x_3 ,t_3) \psi_k^*(\bm x_3 ,t_3) \psi_{k+q}(\bm x_3 ,t_3)
\end{eqnarray}
with $\bm x_3=\hbar \bm q t_3/m$. We approximate the laser beam geometry as rectangular with
\begin{eqnarray}
\label{rectangular}
\bm q = k_s
\left(\begin{array}{c}
1 \\ 1 \\ 0
\end{array}
\right)
,\qquad
\bm k = k_s
\left(\begin{array}{c}
1 \\ -1 \\ 0
\end{array}
\right)
.\end{eqnarray}
This yields
\begin{eqnarray}
\label{int}
\lefteqn{
\psi_q\left(\frac{\hbar \bm q}m t ,t \right)
}\nonumber \\
&=& \frac{1}{i\tau} \sqrt{\frac{21}{8\mathcal V}}
\int_0^{t_2(t)} dt_3 \left(1-\frac{t_3^2}{t_1^2}\right) \sqrt{1-\frac{t_3^2}{t_0^2}}
,\end{eqnarray}
where we introduced two versions of the timescale which describes saturation
\begin{eqnarray}
\label{t0-t1}
t_0 = \frac{m R_y}{2\hbar k_s}
,\qquad
t_1 = \frac{2t_0}{\sqrt{1+\epsilon^2}}
.\end{eqnarray}
$t_0$ and $t_1$ are connected by the aspect ratio $\epsilon=R_y/R_x$. The measured trap frequencies yield $\epsilon=\omega_x/\omega_y=70 \mbox{ Hz}/20 \mbox{ Hz}=3.5$. Eq.\ \eqref{t0-t1} predicts $t_0=2.3$ ms.

The condition that Eq.\ \eqref{u} will be used only if it is real, translates to the upper bound of the integral
\begin{eqnarray}
t_2(t)=\min\{t,t_0,t_1\}
.\end{eqnarray}
The integral has the analytic solution
\begin{eqnarray}
\psi_q\left(\frac{\hbar \bm q}m t ,t \right)
= \frac{t_0}{i\tau} \sqrt{\frac{21}{8\mathcal V}} f\left(\frac{t_2(t)}{t_0}\right)
\end{eqnarray}
with
\begin{eqnarray}
f(x)
&=&
\frac{1}{32} \Big( x \sqrt{1-x^2} [17+\epsilon^2-2x^2(1+\epsilon^2)]
\nonumber
\\ &&
+(15-\epsilon^2) \arcsin x \Big)
.\end{eqnarray}
Rather than performing a similar calculation for each trajectory and eventually carrying out a spatial integral, we assume that this trajectory is sufficiently representative for the whole wave packet. This yields our second coarse approximation
\begin{eqnarray}
N_q(t)
\propto
\left|\psi_q\left(\frac{\hbar \bm q}m t ,t \right)\right|^2
.\end{eqnarray}
The factor of proportionality is obtained by matching the behavior at short times with Eq.\ \eqref{N-q}. Using $f(x)=x+\mathcal O(x^3)$, this yields
\begin{eqnarray}
\label{N-q-fit}
N_q(t)
= \frac{t_0^2}{\tau^2} \left|f\left(\frac{t_2(t)}{t_0}\right)\right|^2
.\end{eqnarray}
Finally, we add an empiric overall offset $N_\mathrm{offset}$ to this expression for $N_q(t)$. Hence, our model has three free fit parameters: $\tau$, $t_0$, and $N_\mathrm{offset}$.

A fit of this model to the data in Fig.\ \ref{fig-absorption} yields the best-fit values
$\tau = 2.5$ $\mu$s and $t_0=1.3$ ms. We find good agreement with the prediction $\tau=2.1$ $\mu$s. The coarse approximations used to derive the fitting function are most likely the reason why the agreement with the prediction $t_0=2.3$ ms is not better.

Both fit parameters have a simple graphical interpretation in Fig.\ \ref{fig-absorption}. First, the fit curve reaches $N_q(t)-N_q(0)=10^4$ approximately after the time $t=\tau \sqrt{10^4} =0.25$ ms. Second, the fit curve is time independent for $t\geq t_1= 2t_0/\sqrt{1+\epsilon^2}= 0.69$ ms.

\section{Fringe Visibility and Envelope}
\label{supp-sec-visibility}

\subsection{Qualitative Discussion}
\label{supp-sec-qualitative}

The fringe visibility in Fig.\ \ref{fig-oscillation} is expected to be incomplete. To understand this on a qualitative level, consider a hypothetical scenario in which control light during retrieval would be applied for a very short time such that only a small fraction of the light would be retrieved and the atomic populations would be essentially unchanged. In this case, an appropriate choice of the initial atomic populations could easily balance the intensities emitted along the two pathways generating downward propagating light. In this hypothetical scenario, one would ideally expect unit fringe visibility. If such a measurement yielded a fringe visibility much below unity, then one might be able to draw conclusions about the maximum fidelity of a possible extension of the scheme to a quantum-logic gate.

In our experiment, however, we apply control light during retrieval until no more signal light emerges because this strongly improves the signal-to-noise ratio of the data and it still suffices to demonstrate phase coherence in general. As a result, the atomic populations change drastically during retrieval. Specifically, at the end of the retrieval, there are essentially no atoms left in internal state $|2\rangle$. Neglecting spontaneous emission, this implies that all atoms originally in state $|2,\bm k+\bm q\rangle$ were transferred into state $|1,\bm q\rangle$, which had a small population before retrieval. In other words, the population of state $|1,\bm q\rangle$ grows by a large factor during retrieval. The population growth in state $|1,0\rangle$, however, is given by a factor of $\sim 2$ at best, because this state already has a large population before retrieval and its growth comes dominantly from atoms in state $|2,\bm k\rangle$. Hence, the bosonic stimulation factors along the two pathways that contribute in Fig.\ \ref{fig-oscillation} have very different time evolutions. Therefore, the intensities emitted along the two pathways generating downward propagating light cannot be balanced throughout the whole retrieval process. This reduces the fringe visibility in the time-integrated signal substantially. Hence, the fringe visibility measured in Fig.\ \ref{fig-oscillation} does not allow for a trivial extrapolation for the fidelity of a possible extension of our scheme to a quantum-logic gate.

\subsection{Plane-Wave Model}

To show that we have a reasonable understanding of the value of the fringe visibility $V$ measured in Fig.\ \ref{fig-oscillation}, we now develop a simple model. We start by adapting the Hamiltonian in Eq.\ (7) of Ref.\ \cite{Cola:04:S} to our geometry, obtaining
\begin{eqnarray}
H
&=&
\sum_{n=-\infty}^\infty
\hbar ( \omega_{b,n} \hat b_n^\dag \hat b_n + \omega_{c,n} \hat c_n^\dag \hat c_n )
- \sum_{j\in\{d,r\}} \hbar \delta_j \hat a_j^\dag \hat a_j
\nonumber
\\
&+&
\sum_{n=-\infty}^\infty
\left(i \hbar g
(
\hat a_d^\dag \hat c_{n+1}^\dag \hat b_n
+ \hat a_r^\dag \hat c_n^\dag \hat b_n
)+ \mathrm{H.c.} \right)
.
\end{eqnarray}
The operators $\hat b_n^\dag$ and $\hat c_n^\dag$ create bosonic atoms in states $|2,\bm k + n \bm q\rangle$ and $|1,n \bm q\rangle$, respectively. The first terms in $H$ represent the kinetic energies of these atoms. The internal-state energies for atoms at rest are absorbed in an interaction picture. Equation \eqref{rectangular} yields $\omega_{b,n}=2 (n^2+\nolinebreak 1) \omega_r$ and $\omega_{c,n}=2 n^2 \omega_r$, where $\omega_r=E_\mathrm{rec}/\hbar=\hbar k_s^2/2m$.

The operators $\hat a_d^\dag$ and $\hat a_r^\dag$ create signal photons in plane-wave modes propagating downward and rightward, respectively. $\delta_d$ and $\delta_r$ denote the two-photon detunings of the Raman transition. If we approximate all BECs and the control light applied during retrieval as plane waves, then the two emission modes will be fixed by momentum conservation. The terms in $H$ containing $\delta_d$ and $\delta_r$ express the photon energies in the interaction picture.

The terms $\hat a_d^\dag \hat c_{n+1}^\dag \hat b_n$ and $\hat a_r^\dag \hat c_n^\dag \hat b_n$ describe the emission of signal photons propagating downward and rightward, respectively. These processes are accompanied by the transfer of atoms from state $|2,\bm k + n \bm q\rangle$ to states $|1,(n+\nolinebreak 1)\bm q\rangle$ and $|1,n\bm q\rangle$, respectively. The Hermitian conjugate terms describe reabsorption of previously emitted signal photons. The coupling constant $g$ is proportional to the Rabi frequency of the control laser applied during retrieval. Obviously, it suffices to consider $g$ as real because the phase of $g$ can be absorbed by resetting the phases of the operators $\hat a_d$ and $\hat a_r$.

We now consider the dynamics on a mean-field level. To this end, we calculate the Heisenberg equations of motion for the operators $\hat a_j$, $\hat b_n$, and $\hat c_n$. We take the expectation values of these equations, and obtain equations of motion for mean fields $\widetilde a_j(t)=\langle \hat a_j \rangle$ etc. We assume that the expectation values of products of operators factorize, i.e.\ $\langle \hat c_{n+1}^\dag \hat b_n \rangle=\langle \hat c_{n+1}^\dag \rangle\langle\hat b_n \rangle$ etc. In addition, we make a transition to another interaction picture with $\widetilde a_j=\sqrt N a_j \exp(i\delta_j t)$, $\widetilde b_n=\sqrt N b_n \exp(-i\omega_{b,n}t)$, and $\widetilde c_n=\sqrt N c_n \exp(-i\omega_{c,n}t)$, where $N$ is the total atom number. We obtain
\begin{subequations}
\label{dt-abc}
\begin{eqnarray}
\label{dt-ad}
\partial_t a_d
&=&
g_N {\textstyle \sum_n} c_{n+1}^* b_n e^{-i \Delta_{d,n} t}
,\\
\label{dt-ar}
\partial_t a_r
&=&
g_N {\textstyle \sum_n} c_n^* b_n e^{-i \Delta_{r,n} t}
,\\
\partial_t  b_n
&=&
- g_N \left( a_d  c_{n+1} e^{i \Delta_{d,n} t} + a_r  c_n e^{i \Delta_{r,n} t} \right)
,\\
\partial_t c_n
&=&
g_N \left( a_d^*  b_{n-1} e^{-i \Delta_{d,n-1} t} +  a_r^*  b_n e^{-i \Delta_{r,n} t} \right)
\end{eqnarray}
\end{subequations}
with $g_N=g\sqrt N$, $\Delta_{d,n}=\delta_d+\omega_{b,n}-\omega_{c,n+1}=\delta_d-4n\omega_r$, and $\Delta_{r,n}=\delta_r+\omega_{b,n}-\omega_{c,n}=\delta_r+2\omega_r$.

In our experiment, $g_N\gg \omega_r$ and all light is retrieved in a time very short compared to $1/\omega_r$. Combined with the facts that $|\delta_j t|\ll1$ and that significant population in atomic modes exists only for small $n$, we conclude that $|\Delta_{d,n}t|\ll1$ and $|\Delta_{r,n}t|\ll1$ for all relevant $n$. In the following, we therefore approximate all phase factors in Eq.\ \eqref{dt-abc} as unity.

At this point, the model reproduces our qualitative argument from Sec.\ \ref{supp-sec-qualitative}. Consider a hypothetical scenario in which control light during retrieval would be applied for a very short time such that only a small fraction of the light would be retrieved and the atomic populations $b_n$ and $c_n$ would be essentially unchanged. Specifically, we could ignore the build-up of population in atomic momentum components with $|n|>1$. Here, the build-up of the amplitude $a_d$ would be given by
\begin{eqnarray}
\label{early}
\partial_t a_d = g_N (c_0^* b_{-1} + c_1^* b_0)
.\end{eqnarray}
This equation expresses the quantum interference between the two pathways. In this scenario, one would expect that an appropriate choice of the atomic amplitudes should ideally yield perfect fringe visibility. As explained in Sec.\ \ref{supp-sec-qualitative}, our experiment is not performed in this regime of a short retrieval pulse so that the temporal evolution of the atomic populations $b_n$ and $c_n$ during retrieval needs to be taken into account.

\subsection{Irreversibility}

In our experiment, the BEC has a finite spatial extension. As a result, emitted signal light can spatially leave the BEC, making a subsequent reabsorption impossible. This brings an effective irreversibility into the physical process which is not captured by the above plane-wave model. A full numerical model taking spatial wave packets into account is beyond the scope of our present work. Instead, we add damping coefficients $\gamma_d$ and $\gamma_r$ to model this irreversibility. We replace Eq.\ \eqref{dt-abc} by
\begin{subequations}
\label{diff-eq-final}
\begin{eqnarray}
\partial_t a_d
&=&
- {\textstyle \frac12} \gamma_d a_d
+ g_N {\textstyle \sum_n} c_{n+1}^* b_n
,\\
\partial_t a_r
&=&
- {\textstyle \frac12} \gamma_r a_r
+ g_N {\textstyle \sum_n} c_n^* b_n
,\\
\partial_t  b_n
&=&
- g_N \left( a_d  c_{n+1} + a_r c_n \right)
,\\
\partial_t c_n
&=&
g_N \left( a_d^*  b_{n-1} +  a_r^* b_n \right)
.\end{eqnarray}
The total numbers of photons $N_d(t)$ and $N_r(t)$ irreversibly emitted between times 0 and $t$ in each direction are easily calculated using
\begin{eqnarray}
\partial_t N_d
=
N \gamma_d |a_d|^2
,\qquad
\partial_t N_r
=
N \gamma_r |a_r|^2
.\end{eqnarray}
\end{subequations}
The initial conditions are $a_d=a_r=N_d=N_r=0$ and $b_n=c_n=0$ for $|n|>1$. The times needed to travel through the BEC are proportional to the Thomas-Fermi radii in the corresponding direction, so that the ratio of the damping coefficients is determined by the inverse ratio of the corresponding Thomas-Fermi radii $\gamma_r/\gamma_d=R_y/R_x=\epsilon=3.5$. In general, the nonlinear differential equations \eqref{diff-eq-final} can produce fairly complicated non-harmonic oscillations.

As discussed in Sec.\ \ref{supp-sec-intensities}, $g=2\pi \times 0.6$ MHz for our experiment. Combination with $N\sim 10^6$ yields $g_N=2\pi \times 0.6$ GHz.

In the following, we use the simple estimate $\gamma_d=c/R_y \sim 1\times 10^{13}$ s$^{-1}$ with the vacuum speed of light $c$. Note that slow light \cite{fleischhauer:05:S}, i.e.\ a substantial reduction of the group velocity of the signal light due to the presence of the control light, would only be obtained if the population of state $|2\rangle$ were small, which is not the case in our experiment. We obtain $\gamma_d/g_N \sim 3\times 10^3$. This is deeply in the regime $\gamma_d/g_N \gg 1$, where the oscillations are overdamped. Adiabatic elimination of $a_d$ and $a_r$ from the equations (by formally setting $\partial_t a_d=\partial_t a_r=0$) shows that the typical timescale for light to leave the BEC is $\gamma_d/g_N^2\sim 0.7$ $\mu$s. This value agrees fairly well with our experimentally observed timescale of $\sim 1$ $\mu$s. In this regime, the value of $\gamma_d/g_N$ only determines this timescale but has no other effect on the retrieval dynamics. Hence, even if the group velocity of the signal light were reduced, this would not affect the value of the visibility $V$ predicted by the model, as long as one would stay in the regime $\gamma_d/g_N \gg 1$.

The initial conditions are determined by the pulse areas $\vartheta_1$ and $\vartheta_2$ of the two Raman pulses. The state before the first Raman pulse is described by $c_0=1$. We assume that the first Raman pulse is so short that the kinetic energies can be neglected during this pulse duration. The first Raman pulse then converts the state into
\begin{eqnarray}
\left(\begin{array}{c}
c_0 \\ b_0
\end{array}\right)
=
\left(\begin{array}{cc}
C_1 & -S_1 \\ S_1 & C_1
\end{array}\right)
\left(\begin{array}{c}
1\\ 0
\end{array}\right)
=
\left(\begin{array}{c}
C_1 \\ S_1
\end{array}\right)
.\end{eqnarray}
Similarly, the second Raman pulse converts this state into
\begin{subequations}
\begin{eqnarray}
& \displaystyle
\left(\begin{array}{c}
c_0 \\ b_1
\end{array}\right)
=
\left(\begin{array}{cc}
C_2 & -S_2 \\ S_2 & C_2
\end{array}\right)
\left(\begin{array}{c}
C_1 \\ 0
\end{array}\right)
=
\left(\begin{array}{c}
C_1 C_2 \\ C_1 S_2
\end{array}\right)
,
\\
& \displaystyle
\left(\begin{array}{c}
c_{-1} \\ b_0
\end{array}\right)
=
\left(\begin{array}{cc}
C_2 & -S_2 \\ S_2 & C_2
\end{array}\right)
\left(\begin{array}{c}
0 \\ S_1
\end{array}\right)
=
\left(\begin{array}{c}
-S_1 S_2 \\ S_1 C_2
\end{array}\right)
.
\quad
\end{eqnarray}
\end{subequations}
Here we abbreviated $C_k=\cos\frac{\vartheta_k}2$ and $S_k=\sin\frac{\vartheta_k}2$ for $k\in\{1,2\}$. The subsequent FWM builds up population in $b_{-1}$ and $c_1$. In addition, each state accumulates a phase proportional to its kinetic energy. This yields
\begin{subequations}
\label{initial}
\begin{eqnarray}
b_{-1} = -i \beta c_0^* c_{-1} b_0
,
&&
c_{-1} = - e^{i\chi} S_1 S_2
,
\\
b_0 = e^{i\chi} S_1 C_2
,
\hspace*{1.5mm}
&&
c_0 = C_1 C_2
,\\
b_1 = e^{2i\chi} C_1 S_2
,
&&
c_1 = - i\beta b_0^* b_1 c_0
\end{eqnarray}
\end{subequations}
with $\chi=-2\omega_r t_\mathrm{FWM}$ and with a dimensionless parameter $\beta$ that describes how far the FWM has evolved. For short $t_\mathrm{FWM}$, Eq.\ \eqref{N-q} yields $\beta=\sqrt{8/{21}} {g_{12}t_\mathrm{FWM} N}/{\hbar \mathcal V}$. These initial conditions neglect the depletion of the other states caused by FWM because in our experiment $|b_{-1}|^2\ll 1$ and $|c_1|^2\ll 1$.

We numerically solve the differential equations \eqref{diff-eq-final} with the initial conditions \eqref{initial}. To this end, we introduce a momentum cutoff by setting $b_n(t)=c_n(t)=0$ for $|n|>6$.
In addition to $\gamma_r/\gamma_d=3.5$ and $\gamma_d/g_N=3\times 10^3$, we choose $\vartheta_1=\vartheta_2=\pi/2$ and $\beta=\sqrt2$. This yields $|c_1(0)|^2=1/32\sim 0.03$ which is realistic. The numerical calculation shows that some population is transferred to states with $|n|=2$, but population in states with $|n|>2$ is negligible. The calculation of $N_d$ at long times yields an interference pattern as a function of $\chi$. The pattern is to a good approximation sinusoidal. A sinusoidal fit yields a best-fit value of $V=0.61$ for the visibility.

$V$ is fairly sensitive to varying the parameters, e.g.\ $V=0.36$ is obtained for $(\vartheta_1,\vartheta_2,\beta)=(0.35\pi,0.65\pi,\sqrt2)$ as well as for $(\vartheta_1,\vartheta_2,\beta)=(\pi/2,\pi/2,0.3\sqrt2)$. The approximations used in the model as well as the fact that the experimental control over the parameters is not perfect let the experimentally observed value $V=0.35$ from Fig.\ \ref{fig-oscillation} seem reasonable. In particular, the finite visibility in this particular measurement does not indicate experimental problems that would force us to conclude that the fidelity of a possible extension of our scheme to a quantum logic gate would be limited.

\subsection{Envelope of the Interference Pattern}

Now, we develop a simple, analytic model for the envelope of the interference pattern observed in Fig.\ \ref{fig-oscillation}. Downward emission is generated in two pathways, namely from $|2,\bm k-\bm q\rangle$ to $|1,0\rangle$ and from $|2,\bm k\rangle$ to $|1,\bm q\rangle$. We concentrate on the first pathway. To obtain a simple estimate for the envelope, we assume that each atom in state $|2,\bm k-\bm q\rangle$ will emit exactly one photon propagating downward if at the position of this atom there is a nonzero density of atoms in state $|1,0\rangle$. This neglects the competition with rightward emission and the interference with the other pathway.

Hence, we need to calculate the number $N_{k-q}^\mathrm{emit}$ of atoms in $|2,\bm k-\bm q\rangle$ that have spatial overlap with any atomic density in $|1,0\rangle$. For the following integration, we choose the coordinate origin such that the two relevant BECs lie on the $y$ axis at positions $\pm y_0$. $N_{k-q}^\mathrm{emit}$ is given by an integral over the density $n(\bm x)$ of the BEC $|2,\bm k-\bm q\rangle$ centered at $y=-y_0$. We approximate all BECs as Thomas-Fermi parabolae with identical Thomas-Fermi radii. According to Eq.\ \eqref{u} we obtain in the new coordinate system
\begin{eqnarray}
n(\bm x)
= \frac{N_{k-q}}{\mathcal V} \left( 1 -\frac{x^2}{R_x^2} -\frac{(y+y_0)^2}{R_y^2} -\frac{z^2}{R_z^2} \right)
,\end{eqnarray}
where this is positive and zero otherwise. The integration boundaries have to be chosen such that only those atoms are counted that lie within the Thomas-Fermi parabola of the BEC $|1,0\rangle$ centered at $y=y_0$. For $0\leq y_0 \leq R_y$, the integration boundary is set by the BECs at $y=y_0$ and $y=-y_0$ for $y<0$ and $y>0$, respectively. We substitute $x=R_x \rho \cos\varphi$ and $z=R_z \rho \sin\varphi$. We obtain for $0\leq y_0 \leq R_y$
\begin{eqnarray}
\lefteqn{
\frac{N_{k-q}^\mathrm{emit}}{2\pi R_x R_z N_{k-q}}
=
\int_0^{R_y-y_0} dy \int_0^{\sqrt{1-(y+y_0)^2/R_y^2}} d\rho \; \rho n(\bm x)
}
\nonumber\\
&&
+ \int_{-(R_y-y_0)}^0 dy \int_0^{\sqrt{1-(y-y_0)^2/R_y^2}} d\rho \; \rho n(\bm x)
.\qquad\qquad
\end{eqnarray}
This yields for all $y_0$
\begin{eqnarray}
\label{N-emit}
\frac{N_{k-q}^\mathrm{emit}}{N_{k-q}}
= h\left(\frac{y_0}{R_y} \right)
= h\left(\frac{t_\mathrm{FWM}}{2t_0} \right)
,\end{eqnarray}
where we used that the distance between the two relevant BECs amounts to $2y_0=2\hbar k_s t_\mathrm{FWM}/m=R_y t_\mathrm{FWM}/t_0$ with $t_0$ from Eq.\ \eqref{t0-t1}. In addition, we introduced
\begin{eqnarray}
h(x) =
\left\{\begin{array}{ll}
(1-x)^3 (1+3x + x^2), & 0\leq x\leq 1 \\
0, & 1<x \\
\end{array}\right.
\end{eqnarray}
and $h(-x)=h(x)$. Within our approximation, the fraction $h$ of the atoms in $|2,\bm k-\bm q\rangle$ will emit downward. A similar argument for the second pathway yields the same $h$.

As $N_{k-q}$ is given by an expression that is analog to Eq.\ \eqref{N-q-fit}, we obtain a simple estimate for the envelope of the downward emission by multiplying Eqs.\ \eqref{N-q-fit} and \eqref{N-emit}. $t_0$ appears as a fit parameter in both equations. We decouple these two fit parameters by choosing the following fitting function for the data in Fig.\ \ref{fig-oscillation}
\begin{eqnarray}
y_0 + A \left|f\left(\frac{t_2(t)}{t_0}\right)\right|^2 h\left(\frac{t}{2t_4}\right)
\Big(1+V \cos \omega(t-t_5) \Big)
.\end{eqnarray}
Here $y_0$ and $t_5$ are offsets, $A$ is an amplitude, and we expect $t_4=t_0$. A fit of this model yields a curve that looks pretty similar to the curve shown in Fig.\ \ref{fig-oscillation}. The best-fit values $V=0.38$ and $\omega/2\pi=15.4$ kHz agree well with the results from the simple fit shown in Fig.\ \ref{fig-oscillation}. With the best-fit values $t_0=0.97$ ms and $t_4=0.42$ ms, the envelope itself is well approximated by a Gaussian. We attribute the deviation from the prediction $t_0=t_4=2.3$ ms to the coarse approximations used to derive the envelope.

\section{Experimental Details}

\subsection{Rabi Frequencies}
\label{supp-sec-intensities}

The parameters in our experiment varied between different measurements. As an example, we list the parameters for Fig.\ \ref{fig-oscillation}. The applied laser powers are $(P_{s,1},P_{s,2},P_{c,1},P_{c,2},P_{c,3})=(1,1,5,75,180)$ $\mu$W. Here and in the following, the indices $s$ and $c$ refer to signal and control light whereas the indices 1, 2, and 3 refer to the pulse number. The beam waists ($1/e^2$ radii of intensity) are $(w_{s,1},w_{c,1},w_{c,2})=(0.17,0.32,1.8)$ mm. The use of identical beam paths yields $w_{s,1}=w_{s,2}$ and $w_{c,1}=w_{c,3}$. From these parameters, one can estimate intensities and electric field amplitudes according to $I=2P/\pi w^2$ and $\mathcal E=\sqrt{2I/c\epsilon_0}$, where $\epsilon_0$ is the vacuum permittivity. We first consider the closed cycling transition $|F=2,m_F=2\rangle \leftrightarrow \linebreak[1] |F'=3,m_F'=3\rangle$ on the $D_2$ line at $\lambda=780$ nm. Its excited-state decay rate $\Gamma=1/(26\ \rm ns)$ yields a dipole matrix element $d^\mathrm{cyc}=2.5\times10^{-29}$ Cm and a saturation intensity $I_\mathrm{sat}=1.6$ mW/cm$^2$. If the light drove this transition, then the Rabi frequency would be given by $\Omega^\mathrm{cyc}=d^\mathrm{cyc}\mathcal E/\hbar=\Gamma \sqrt{I/2I_\mathrm{sat}}$.

The detunings of the control light from the $|2\rangle\to|1'\rangle$ transition for the three pulses are $(\Delta_1,\Delta_2,\Delta_3)=2\pi\times(406,406,512)$ MHz. The effective two-photon Rabi frequency in our experiment is the superposition of the contributions from the two excited states $|1'\rangle$ and $|2'\rangle$
\begin{eqnarray}
\Omega_\mathrm{eff} = \frac{\Omega_s^\mathrm{cyc} \Omega_c^\mathrm{cyc}}2 \left( \frac{\sqrt3}{12} \frac{1}{\Delta} - \frac{\sqrt3}{12} \frac{1}{\Delta-\Delta_\mathrm{HFS}}\right)
\end{eqnarray}
with the excited-state hyperfine splitting $\Delta_\mathrm{HFS}/2\pi=812$ MHz on the $D_1$ line at $\lambda=795$ nm. The factors $\pm\sqrt3/12$ represent the products of the dipole matrix elements of the involved transitions in units of $d^\mathrm{cyc}$.

With pulse durations of $(t_1,t_2)=(23,35)$ $\mu$s, we estimate pulse areas $\vartheta=\Omega_\mathrm{eff}t$ of $(\vartheta_1,\vartheta_2)=\pi\times (0.49,0.51)$ which agree fairly well with the observed atomic population transfer.

For retrieval, the parameter $g$ from Ref.\ \cite{Cola:04:S} corresponds to $\Omega_\mathrm{eff}$ with the electric field amplitude $\mathcal E_{s,3}=\sqrt{\hbar\omega/2\epsilon_0 \mathcal V_q}$, the quantization volume $\mathcal V_q$, and $\omega=2\pi c/\lambda$. The connection between a homogeneous model with $\mathcal V_q$ and a Thomas-Fermi parabola is made by approximating the quantization volume $\mathcal V_q$ by the BEC volume $\mathcal V$ from Eq.\ \eqref{V}. This is because the relevant physical quantity is the typical atomic density which equals $N/\mathcal V_q$ and $N/\mathcal V$, respectively. We obtain $g=\Omega_{\mathrm{eff},3} = 2\pi\times 0.6$ MHz.

\subsection{Raman AMW during the second Raman pulse}

During the second Raman pulse, Raman AMW occurs as a side effect with atoms transferred from $|2,\bm k\rangle$ to $|1,0\rangle$ under emission of downward propagating light. The effect on the atomic populations is negligible. But the emitted light reaches our EMCCD camera which needs $\sim 0.1$ ms to erase this signal from the relevant pixels. This is why no data are shown for shorter times in Fig.\ \ref{fig-oscillation}.

\subsection{No Raman AMW during the depletion pulse}

The depletion pulse propagates downward. After the corresponding absorption recoil, there is no emission direction available such that Raman AMW could occur. Hence, the depletion pulse causes only spontaneous light scattering. The depletion pulse depletes the $F=2$ states and operates on an open transition so that an appreciable fraction of the atoms scatters only one photon. This does not suffice to spatially remove the atoms quickly from the BEC. Nevertheless, the spontaneous scattering aborts the FWM, because FWM occurs with BECs, but not with atoms incoherently distributed over a large number of momentum states.

\end{document}